\documentclass[twocolumn,english,reprint,floatfix,amsmath,amssymb,aps,prl,superscriptaddress]{revtex4-1}
\usepackage{XCharter}
\usepackage[utf8]{luainputenc}
\usepackage{geometry}
\usepackage{mathrsfs}
\usepackage{amsmath}
\usepackage{amssymb}
\usepackage{graphicx}

\makeatletter
\usepackage{dcolumn}\usepackage{bm}\usepackage{color}\usepackage{comment}\usepackage{url}\usepackage{XCharter}\usepackage{xspace}\usepackage{xr}\usepackage{cleveref}

\usepackage{babel}

\makeatother

\usepackage{babel}
\begin{document}
\title{Moment analysis of two-dimensional active Brownian run-and-tumble
particles}
\author{Aoran Sun}
\email{Corresponding author. sunaoran16@mails.ucas.ac.cn}

\affiliation{Beijing National Laboratory for Condensed Matter Physics, Institute
of Physics, Chinese Academy of Sciences, Beijing 100190, China}
\affiliation{School of Physical Sciences, University of Chinese Academy of Sciences,
Beijing 100049, China}
\author{Da Wei}
\affiliation{Beijing National Laboratory for Condensed Matter Physics, Institute
of Physics, Chinese Academy of Sciences, Beijing 100190, China}
\author{Yiyu Zhang}
\affiliation{Beijing National Laboratory for Condensed Matter Physics, Institute
of Physics, Chinese Academy of Sciences, Beijing 100190, China}
\affiliation{School of Physical Sciences, University of Chinese Academy of Sciences,
Beijing 100049, China}
\author{Fangfu Ye}
\email{Corresponding author. fye@iphy.ac.cn}

\affiliation{Beijing National Laboratory for Condensed Matter Physics, Institute
of Physics, Chinese Academy of Sciences, Beijing 100190, China}
\affiliation{School of Physical Sciences, University of Chinese Academy of Sciences,
Beijing 100049, China}
\affiliation{Wenzhou Institute, University of Chinese Academy of Sciences, Wenzhou,
Zhejiang 325001, China}
\author{Rudolf Podgornik}
\email{Deceased}

\affiliation{School of Physical Sciences, University of Chinese Academy of Sciences,
Beijing 100049, China}
\affiliation{Beijing National Laboratory for Condensed Matter Physics, Institute
of Physics, Chinese Academy of Sciences, Beijing 100190, China}
\affiliation{Wenzhou Institute, University of Chinese Academy of Sciences, Wenzhou,
Zhejiang 325001, China}
\affiliation{Kavli Institute for Theoretical Sciences, University of Chinese Academy
of Sciences, Beijing 100049, China}
\begin{abstract}
We study an {\em active Brownian run-and-tumble particle (ABRTP)}
model, that consists of an active Brownian {\em run state} during
which the active velocity of the particle diffuses on the unit circle,
and a {\em tumble state} during which the active velocity is zero,
both with exponentially distributed time. Additionally we add a harmonic
trap as an external potential. In the appropriate limits the ABRTP
model reduces either to the active Brownian particle model, or the
run-and-tumble particle model. Using the method of direct integration
the equation of motion, pioneered by Kac, we obtain exact moments
for the Laplace transform of the time dependent ABRTP, in the presence
or absence of a harmonic trap. In addition we estimate the distribution
moments with the help of the Chebyshev polynomials. Our results are
in excellent agreement with the experiments. 
\end{abstract}
\maketitle

\section{Introduction}

Active particles differ from Brownian particles in that their motion
can be fueled by the environment \citep{2013_HydrodynamicsofSoftActiveMatter,2016_ActiveParticlesinComplexandCrowdedEnvironments,2022_IrreversibilityandBiasedEnsemblesinActiveMatterInsightsfromStochasticThermodynamics}.
They are quite common in nature and appear in many contexts, such
as molecular motors \citep{2006_ActiveGelsDynamicsofPatterningandSelfOrganization,2007_NonequilibriumMechanicsofActiveCytoskeletalNetworks},
cells \citep{2004_E.ColiinMotion,2008_OutofEquilibriumMicrorheologyinsideLivingCells},
granular materials \citep{2017_NoiseandDiffusionofaVibratedSelfPropelledGranularParticle},
active gels \citep{2010_TheMechanicsandStatisticsofActiveMatter,2019_SignaturesofMotorSusceptibilitytoForcesintheDynamicsofaTracerParticleinanActiveGel},
large (compared with cells) animals \citep{2004_AModeloftheFormationofFishSchoolsandMigrationsofFish,2005_HydrodynamicsandPhasesofFlocks,2014_FlockingataDistanceinActiveGranularMatter,Du_2021},
\textsl{etc.}. Additionally, they can be designed and fabricated with
robot-like qualities \citep{2014_GravitaxisofAsymmetricSelfPropelledColloidalParticles,2016_AcousticTrappingofActiveMatter,2016_ActiveParticlesinComplexandCrowdedEnvironments}.

Besides the common appearance, active particles - even in the case
of a single particle in free space - are fascinating also from a purely
theoretical point of view, as they represent a paradigmatic model
of nonequilibrium, non-Boltzmann statistics \citep{2014_GravitaxisofAsymmetricSelfPropelledColloidalParticles,2015_ActiveBrownianParticlesandRunandTumbleParticlesaComparativeStudy,2016_AcousticTrappingofActiveMatter}.
Two of the most widely discussed minimal theoretical model of active
particles are the {\em active Brownian particle (ABP)} model and
the {\em run-and-tumble particle (RTP)} model. The ABP model can
successfully model (up to a certain accuracy) many known microswimmers
\citep{2012_ActiveBrownianParticlesfromIndividualtoCollectiveStochasticDynamics,2016_ActiveParticlesinComplexandCrowdedEnvironments},
while the RTP mimics the actual motion of certain types of bacteria
\textsl{e.g}., Escherichia coli \citep{2004_E.ColiinMotion,2022_WrappedUptheMotilityofPolarlyFlagellatedBacteria}.
In both models, the velocity of the active particle has a fixed magnitude,
and the direction of the velocity changes stochastically with time.
In the ABP model the velocity diffuses either on a circle or a sphere,
whereas in the RTP model the direction of velocity remains the same
for an exponentially distributed time - (\textsl{the run state})-
and then randomly changes - (\textsl{the tumble state}) - to a different,
randomly chosen direction. \textsl{i.e.}, to another run state. While
apparently simple on first sight, such models already contain a rich
variety of features and can be non-trivial to analyze \citep{2015_ActiveBrownianParticlesandRunandTumbleParticlesaComparativeStudy,2020_RunandTumbleParticlesinTwoDimensionsMarginalPositionDistributions,2020_UniversalSurvivalProbabilityforadDimensionalRunandTumbleParticle}.

For the problem of a single RTP, the time dependent distribution has
been found for the general case (in terms of its Fourier-Laplace transform)
\citep{2013_AveragedRunandTumbleWalks}. Other interesting quantities,
such as the first passage time \citep{2019_NoncrossingRunandTumbleParticlesonaLine,2020_UniversalSurvivalProbabilityforadDimensionalRunandTumbleParticle},
the survival probability \citep{2020_UniversalSurvivalProbabilityforadDimensionalRunandTumbleParticle},
and the distribution of the time of the maximum \citep{2019_GeneralisedArcsineLawsforRunandTumbleParticleinOneDimension},
have also been calculated. The problem of a single ABP appears to
be harder to analyze, and all known analytic distributions seem to
be amenable to complicated infinite series \citep{2015_SmoluchowskiDiffusionEquationforActiveBrownianSwimmers,2016_IntermediateScatteringFunctionofanAnisotropicActiveBrownianParticle}. 

An active particle in an external potential is a natural and interesting
generalization of this problem \citep{2015_RunandTumbleDynamicsofSelfPropelledParticlesinConfinement,2020_VelocityandDiffusionConstantofanActiveParticleinaOneDimensionalForceField,2024_ConfinedRunandTumbleParticleswithNonMarkovianTumblingStatistics},
since the active particle may eventually reach a non-Boltzmann, nonequilibrium
steady state \citep{2022_ExactPositionDistributionofaHarmonicallyConfinedRunandTumbleParticleinTwoDimensions,2022_PositingtheProblemofStationaryDistributionsofActiveParticlesAsThirdOrderDifferentialEquation,2023_RunandTumbleOscillatorMomentAnalysisofStationaryDistributions,2024_ActiveOscillatorRecurrenceRelationApproach}.
For the special case of a harmonic potential, which models the optical
\citep{1997_OpticalTrappingandManipulationofNeutralParticlesUsingLasers,2022_Activecolloidsinharmonicopticalpotentials}
or the acoustic \citep{2016_AcousticTrappingofActiveMatter} tweezer
trap common in experiments, the exact steady state distribution for
the RTP in 1D \citep{2015_PressureIsNotaStateFunctionforGenericActivefluids,2019_RunandTumbleParticleinOneDimensionalConfiningPotentialsSteadyStateRelaxationandFirstPassageProperties}
and 2D \citep{2022_PositingtheProblemofStationaryDistributionsofActiveParticlesAsThirdOrderDifferentialEquation,2023_RunandTumbleOscillatorMomentAnalysisofStationaryDistributions,2024_ActiveOscillatorRecurrenceRelationApproach}
has been found, as well as the moments of the steady state distribution
in the 3D case \citep{2022_PositingtheProblemofStationaryDistributionsofActiveParticlesAsThirdOrderDifferentialEquation,2023_RunandTumbleOscillatorMomentAnalysisofStationaryDistributions,2024_ActiveOscillatorRecurrenceRelationApproach}.
For the ABP, the situation is again worse, as the steady state distribution
is found only in the form of a infinite series \citep{2020_SteadyStateofanActiveBrownianParticleinaTwoDimensionalHarmonicTrap,2022_AnalyticSolutionofanActiveBrownianParticleinaHarmonicWell},
though the exact steady state moments can be calculated recursively
\citep{2024_ActiveOscillatorRecurrenceRelationApproach}.

In general, the theoretical analysis of these models, especially in
the presence of an external field, can be very challenging. Besides
a few known exact solutions and some examples of perturbation analysis
\citep{2020_TowardtheFullShortTimeStatisticsofanActiveBrownianParticleonthePlane,2021_DirectionReversingActiveBrownianParticleinaHarmonicPotential,2023_NonequilibriumSteadyStateofTrappedActiveParticles},
many theoretical problems still seem to be beyond what is currently
feasible.

Most of the theoretical works focus on the Fokker-Planck equation.
However, it is usually highly challenging to solve. Recently, the
method of direct integration of the correlation functions, pioneered
by Kac \citep{1974_AStochasticModelRelatedtotheTelegraphersEquation},
has been applied to both the RTP and ABP models, to calculate time
dependent moments, with or without the background provided by the
harmonic trap \citep{2025_ExactMomentsforaRunandTumbleParticlewithaFiniteTumbleTimeinaHarmonicTrap}.
Furthermore, this method allows for the inclusion of a tumble state,
with zero active velocity and exponentially distributed time, into
the RTP model, without any major complication, and the exact steady
state distribution in 1D can be obtained in this way.

\begin{figure}[t]
\includegraphics[width=0.4\textwidth]{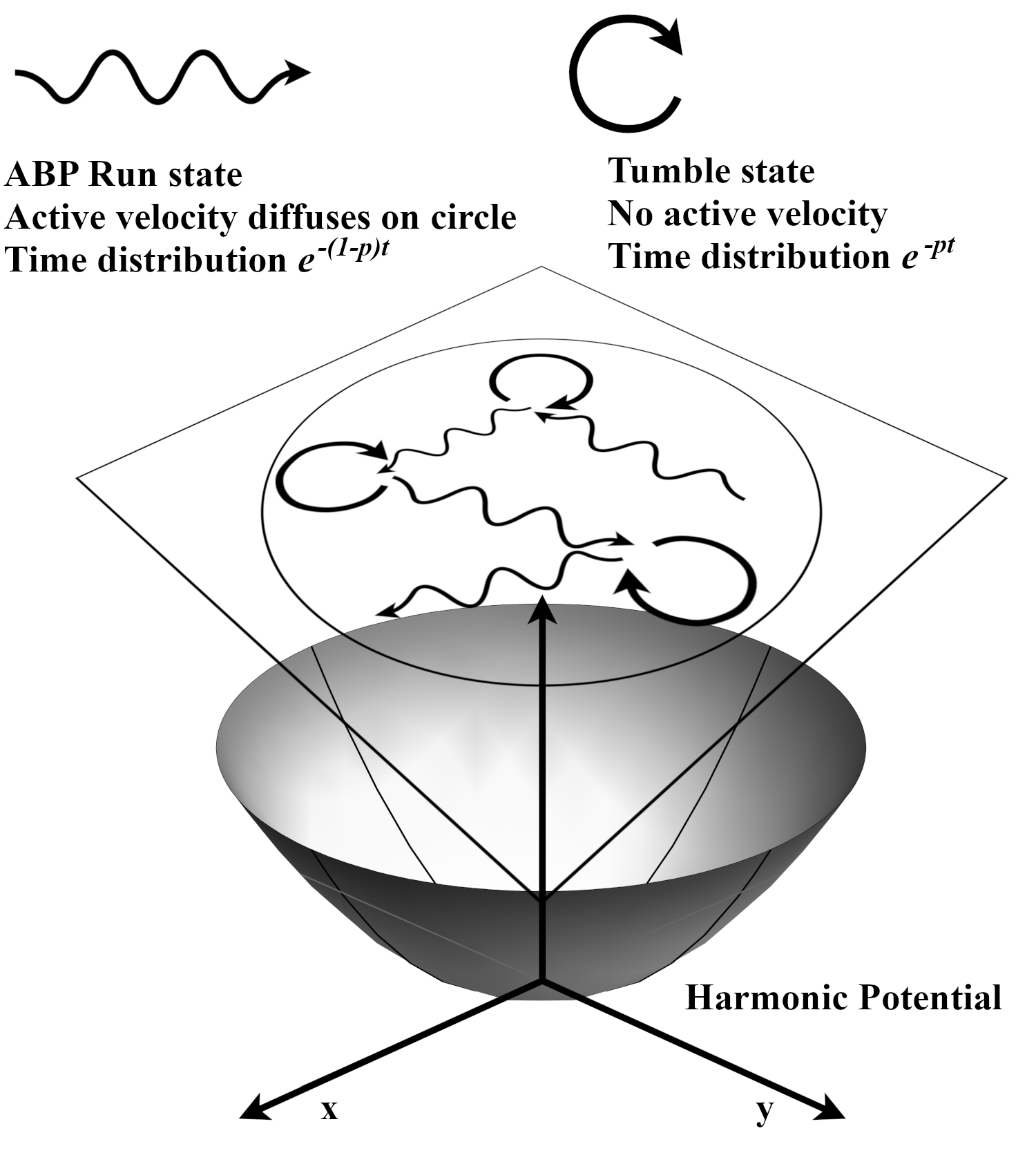}\caption{Schematic drawing of the 2D ABRTP model. The velocity of the particle
is a consequence of two components, the pull, $-br$, of harmonic
potential $b\left|r\right|^{2}$, depending only on the position,
and the active velocity, which switches randomly between zero (\textsl{the
tumble state}), and a vector of magnitude $v$ (\textsl{the run state}).
In the run state, the active component will diffuse along a circle
from a randomly chosen starting point at the beginning of the run
state. The time between these switches is exponentially distributed,
with rate $\left(1-p\right)\gamma$ and $p\gamma$, respectively for
the run and the tumble state, and each choice of the active velocity
at the beginning of each run state is independent.}
\label{fig:ad} 
\end{figure}

Equipped with this methodological tool, we now focus on a better description
of the active particles by proposing and studying the (2D) active
Brownian run-and-tumble particle (ABRTP) model. Invoking the example
of the motion of Escherichia coli \citep{2004_E.ColiinMotion}, it
is clear that, in the run state, the cell moves in a line that is
not entirely straight. Furthermore, experiments with, \textsl{e.g}.,
the BV2 cells, show not only the run-and-tumble pattern, but also
a significant angular diffusion of the velocity during the run state
\citep{2023_RunandTumbleDynamicsandMechanotaxisDiscoveredinMicroglialMigration}.
We therefore propose to modify the RTP model, by replacing all the
run states with active Brownian motion, thus resulting in the ABRTP
model, in which - just like in the RTP model - both the run and the
tumble states feature with an exponentially distributed time. At the
beginning of the run state, an active velocity is randomly chosen,
but unlike in the RTP model, where such active velocity will remain
the same until the end of the run state, in the ABRTP model, the active
velocity will diffuse on a circle. The tumble state is the same as
in the RTP, and the particle has zero active velocity. Subsequently,
in the new run state, the active velocity will start a new diffusion
from a new randomly chosen point on the circle, thus repeating the
ABRTP algorithm. See Fig. \ref{fig:ad} for a schematic presentation
of the model.

The ABRTP model naturally reduces to either the ABP or the RTP under
the appropriate limit. But apart from that, there seems to be no related
theoretical results to compare with. To validate the new model, we
compare with the available experimental data of the BV2 cells. The
most direct result from the theoretical calculation are the moments,
yet the high order moments will be dominated by the few events of
very large variables, thus reducing significantly the effective dataset,
whereas being effectively ignorance about distributions of small data,
rendering the comparison unreliable except for a few very low order
ones \textsl{e.g}., the mean squared displacement (MSD). We therefore
propose to check our results by comparing the expectation of the Chebyshev
polynomials, which can be reliably calculated both analytically from
the exact moments, and numerically from data. We will show that the
agreement is quite satisfactory.

The rest of the paper is organized as follows. In Section \ref{sec:Kac's-method-revisited}
we discuss the methodology of Kac. 
In Subsection \ref{subsec:Active-Brownian-particles-1} we briefly
demonstrate this methodology on the ABP problem and in the Subsection
\ref{subsec:Run-and-tumble} on the RTP, since the ideas and results
will be directly referred to later. In Section \ref{sec:Active-Brownian-run}
we present the ABRTP model and apply the Kac method, obtaining a programmable
method to calculate the Laplace transform of time depending moments.
In Subsection \ref{subsec:Equation-of-motion} we present the equation
of motion to be used for our model, and in Subsection \ref{subsec:Diagram-laws}
we apply the Kac method to calculate the moments and summarize the
results in the form of diagram laws. In Subsection \ref{subsec:Recursive-relation}
we derive from the diagram law a Volterra difference equation that
is more easily implemented on computers. The Volterra equation is
very similar to the RTP model, but differs in the form of the contribution
from each ABP run state, which can be calculated from another programmable
recursive relation obtained in Subsection \ref{subsec:Recursive-relation-for},
similar to the case studied in \citep{2024_ActiveOscillatorRecurrenceRelationApproach}.
In Section \ref{sec:Results} we list some of the results. In Subsection
\ref{subsec:Free-space} we compare our model against experimental
data available for BV2 cells \citep{2023_RunandTumbleDynamicsandMechanotaxisDiscoveredinMicroglialMigration},
by comparing the expectation of the Chebyshev polynomials. In Subsection
\ref{subsec:Steady-state-in} we consider the steady state within
a harmonic potential, where we examine the scaling of the high order
moments and infer the distribution from its moments with the help
of the Chebyshev polynomials. Finally in \ref{sec:Conclusion} we
state the conclusions with a brief summary and some speculation regarding
the unsolved problems.

\section{Kac's method revisited }

\label{sec:Kac's-method-revisited}

The Kac method starts with the Langevin equation of motion for an
overdamped particle in a harmonic potential: 
\begin{equation}
\dot{x}\left(t\right)=-bx\left(t\right)+F\left(t\right),\label{eq:general=00003D00003D00003D00003D00003D00003D000020equation=00003D00003D00003D00003D00003D00003D000020of=00003D00003D00003D00003D00003D00003D000020motion}
\end{equation}
where $x$ is (a single component of) the position of the particle,
$b$ is the strength of the harmonic potential, $F$ is the time dependent
force that likely contains the stochastic active velocity. Here we
will assume spherical symmetry, and focus on the distribution instead
of the correlation, therefore a focus on a single coordinate component
is sufficient. Eq. \ref{eq:general=00003D00003D00003D00003D00003D00003D000020equation=00003D00003D00003D00003D00003D00003D000020of=00003D00003D00003D00003D00003D00003D000020motion}
thus appears as a scalar equation in 1D, even though our problem is
actually in 2D. The dimensionality will came in through $F$, which
is the result of the projection to the coordinate axis under consideration.

For fairly general forms of $F$, the equation of motion can be solved
at least formally as: 
\begin{align}
x\left(t\right) & =x\left(0\right)e^{-bt}+\int_{0}^{t}F\left(t\right)e^{-b\left(t-s\right)}ds.\label{eq:SolutionGeneralODE}
\end{align}
For simplicity we have assumed $x\left(0\right)=0$. We then seek
to calculate the moments of the time dependent distribution: 
\begin{align}
\left\langle x\left(t\right)^{l}\right\rangle = & ~l!\int_{0\leqslant t_{1}\leqslant...\leqslant t_{l}\leqslant t}\!\!\!\!\!\!\!\!\!\!\!\!\!\!\!\prod_{k}dt_{k}e^{-b\left(lt-\sum_{k}t_{k}\right)}\left\langle \prod_{k}F\left(t_{k}\right)\right\rangle ,\label{eq:MomentsGeneral}
\end{align}
where we have used the symmetry properties to re-order the terms according
to their time parameter. While the integral might appear formidable
on first sight, it can be rearranged into a convolution, and the Laplace
transform of the integral can be calculated easily and systematically.
Indeed, the exponential term $e^{-b\left(lt-\sum_{k}t_{k}\right)}$,
associated with the harmonic potential, can be rearranged using the
identity: 
\begin{equation}
lt-\sum_{k=1}^{l}t_{k}=\sum_{k=1}^{l}k\left(t_{k+1}-t_{k}\right),\label{eq:integraltoconvolution}
\end{equation}
where we have used the convention $t_{l+1}=t$. Depending on the details
of the model, if the correlation function of $F$ can also be written
as a product of $l$ functions, such that the $k$-th depends on $t_{k}-t_{k-1}$
only (where the convention is $t_{0}=0$), then the integral is a
convolution of $l$ functions, and its Laplace transform can be calculated
by multiplying the Laplace transforms of $l$ individual functions.

Once we have calculated the Laplace transform of the moments, we can
obtain the exact distribution by considering the Laplace transformed
characteristic function $\varphi\left(\omega,\xi\right)=\mathscr{L}\left(\left\langle e^{i\omega x}\right\rangle \right)\left(\xi\right)$,
where $\mathscr{L}$ represents the Laplace transform, and $\xi$
is the variable of the Laplace transform. The RHS can be calculated
by expanding the exponential and exchanging the order of summation
and averaging. On the other hand, the characteristic function is the
Fourier-Laplace transform of the distribution function, and the distribution
function can be obtained by the inverse transforms. If only the steady
state distribution is needed, then the inverse Laplace transform can
be obtained more straightforwardly, by using the simple identity:
\begin{equation}
\lim_{t\rightarrow\infty}f\left(t\right)=\lim_{\xi\rightarrow0}\xi\mathscr{L}\left(f\right)\left(\xi\right).
\end{equation}
If there are difficulties in any of these steps, such that the exact
distribution function cannot be obtained, there is still much to learn
about the distribution based solely on its moments. This reverberates
with the century-old problem in mathematics referred to as {\em
the moments problem} \citep{2020_TheClassicalMomentProblemandSomeRelatedQuestionsinAnalysis}.
For example, asymptotically decreasing moments (after appropriately
rescaling them) imply that the distribution is compactly supported
within $\left[-1,1\right]$. In this case, a power scaling law: 
\begin{equation}
\left\langle \left(x\right)^{l}\right\rangle \propto l^{-\alpha}
\end{equation}
can be translated into an explicit form of the distribution function
near the boundary: 
\begin{equation}
p\left(x\right)\propto\left(1-\left|x\right|\right)^{\alpha-1}.
\end{equation}
Furthermore, it is possible to numerically approximate the expectation
of any smooth function $f$ supported within $\left[-1,1\right]$,
by expanding $f$ using Chebyshev polynomials, and exchange the order
of summation and averaging, such that: 
\begin{equation}
\left\langle f\right\rangle \thickapprox\sum_{n=0}^{N}\sum_{i=0}^{N}\frac{2}{p_{i}p_{n}N}\cos\left(\frac{in\pi}{N}\right)f\left(x_{i}\right)\left\langle T_{n}\right\rangle ,
\end{equation}
where $x_{i}=\cos\frac{\pi i}{N}$;$p_{i}=2$ if $i=0$ or $i=N$,
and $p_{i}=1$ otherwise, and $T_{n}$ stands for the Chebyshev polynomials
(of the first kind), defined by \citep{2013_ChebyshevandFourierSpectralMethodsSecondRevisedEdition,2014_SolvingTranscendentalEquations}:
\begin{equation}
T_{n}\left(\cos\theta\right)=\cos n\theta.
\end{equation}
This, in turn, leads to an estimation of the distribution density
$p$ itself, at specific points: 
\begin{equation}
p\left(\cos\frac{\pi j}{N}\right)\sim\sum_{n=0}^{N}\frac{2\cos\left(\frac{in\pi}{N}\right)}{p_{n}\pi\sin\left(\frac{i\pi}{N}\right)}\left\langle T_{n}\right\rangle .
\end{equation}
In the rest of this section, we briefly demonstrate how the Kac method
can be applied to two most common models of active particles, the
ABP and the RTP, not only for pedagogical reasons, but also because
some of the results will be needed for the ABRTP later on.

\subsection{Active Brownian particles}

\label{subsec:Active-Brownian-particles-1}

In 2D the ABP is defined by positing: 
\begin{eqnarray}
F\left(t\right) & = & v\cos\theta\left(t\right)=\frac{v}{2}\sum_{a=\pm1}e^{ia\theta\left(t\right)}\\
\dot{\theta}\left(t\right) & = & \sqrt{2D}\eta\left(t\right),
\end{eqnarray}
where $v$ is the magnitude of the active velocity, $\theta$ is the
angle representing the direction of the active velocity, $D$ is the
diffusion constant and $\eta$ the standard Gaussian white noise.

Perhaps the most straightforward way to evaluate the correlation function
is to use the identity valid for Gaussian variables: 
\[
\left\langle e^{A}\right\rangle =e^{\left\langle A\right\rangle +\frac{1}{2}\left(\left\langle A^{2}\right\rangle -\left\langle A\right\rangle ^{2}\right)},
\]
and for simplicity we shall let the initial condition for $\theta$
be $\theta\left(0\right)=0$, and therefore $\left\langle \theta\left(t_{i}\right)\theta\left(t_{j}\right)\right\rangle =2D\min\left(t_{i},t_{j}\right)$.
One can then prove that 
\begin{equation}
\left\langle \prod_{k}F\left(t_{k}\right)\right\rangle =\frac{v^{l}}{2^{l}}\sum_{a_{i}=\pm1}e^{-D\sum_{k=1}^{2l-1}\left(t_{k+1}-t_{k}\right)\left(\sum_{i=k+1}^{2l}a_{i}\right)^{2}},
\end{equation}
and thus, by using Eq. \ref{eq:MomentsGeneral}, we finally arrive
at 
\begin{equation}
\mathscr{L}\left(\left\langle x^{l}\right\rangle \right)\left(\xi\right)=\frac{v^{l}l!}{2^{l}}\sum_{a_{i}=\pm1}\prod_{k=0}^{l}\frac{1}{D\left(\sum_{i=k+1}^{l}a_{i}\right)^{2}+bk+\xi}.
\end{equation}
Here it is understood that $\sum_{i=l+1}^{l}...=0$. The steady state
moments then agree exactly with the examples given in \citep{2024_ActiveOscillatorRecurrenceRelationApproach}.

A simple yet important special case is $l=1$. In this case, we need
to multiply together the terms with $k=0$ and $k=1$. If $k=0$,
then $\sum_{i=1}^{1}a_{i}=a_{1}$, and the term to multiply will be
$1/\left(Da_{1}^{2}+\xi\right)$; if $k=1$, then by convention $\sum_{i=2}^{1}a_{i}=0$
and we are left with $1/\left(b+\xi\right)$. We then multiply them
together, sum over $a_{1}=\pm1$ and multiply $v/2$, obtaining:
\begin{equation}
\mathscr{L}\left(\left\langle x\right\rangle \right)\left(\xi\right)=\frac{v}{\left(D+\xi\right)\left(b+\xi\right)},
\end{equation}
which links the model parameter $v$ to experimentally measured averaged
velocity $v_{e}$. In most experiments, velocities are measured by
measuring the changes of position between a time interval $\Delta t$,
then average over measurements, thus $v_{e}$ corresponds, at least
for reasonably small $\Delta t$, to $\left\langle x\left(\Delta t\right)/\Delta t\right\rangle $
After the inverse Laplace transform, it can be shown that: 
\begin{equation}
v_{e}=\left\langle \frac{x\left(\Delta t\right)}{\Delta t}\right\rangle =v\frac{e^{-b\Delta t}-e^{-D\Delta t}}{\Delta t\left(D-b\right)}.\label{eq:VeABP}
\end{equation}
Eq. \ref{eq:VeABP} will be used for the experimental check in Subsection
\ref{subsec:Free-space}.

\subsection{Run-and-tumble particles}

\label{subsec:Run-and-tumble}

Detailed study of the RTP problem with finite tumble time using the
Kac method has been published elsewhere \citet{2025_ExactMomentsforaRunandTumbleParticlewithaFiniteTumbleTimeinaHarmonicTrap},
and here we only list the most important ideas and results that would
be necessary for the analysis of the ABRTP model.

Briefly, we may write $F\left(t\right)=F\left(K\left(t\right)\right)$
for the stochastic active velocity, where $K$ stochastically flips
between $R$ and $T$, and the life time of each state is exponentially
distributed, with rate $\gamma_{R}$ and $\gamma_{T}$ respectively
for $R$ and $T$. $F$ is then determined from $K$ by $F\left(T\right)=0$,
and $F\left(R\right)$ randomly chosen, according to dimension of
the problem, from two points $\left(\pm v\right)$, a circle or a
sphere, then projected into 1D. 

To study this model we apply the law of total expectations \citep{2013_StochasticProcesses},
or a case by case discussion:
\begin{equation}
\left\langle \cdot\right\rangle =\sum_{diag}\left\langle \cdot|diag\right\rangle \mathbb{P}\left(diag\right),
\end{equation}
here we require $diag$ to represents different cases under consideration
and $\mathbb{P}\left(diag\right)$ the probability that case $diag$
happens. It is easy to see that for conditional expectation $\left\langle \cdot|diag\right\rangle $
to be non-zero, the particle must be in the run state at all times
appearing in the correlation function. Thus, for any two times, either
the particle is in the same run state, and the active velocity remains
the same, or the particle has enter the tumble state at least once,
and the active velocities are no longer correlated. These possibilities
of the state that the particle is in at the time appearing in the
correlation function form different cases we need to consider, each
of them may be represent as a diagram: we use the dots on a line to
represent the times in the correlation function, and connect the dots
with solid line if the particle is in the same run state at the times
represented by these dots. \textsl{E.g.}, the correlation function
of order eight contains the following diagram: 
\begin{eqnarray}
\bullet_{t_{8}}-\bullet_{t_{7}}-\bullet_{t_{6}}-\bullet_{t_{5}}\quad\bullet_{t_{4}}-\bullet_{t_{3}}\quad\bullet_{t_{2}}-\bullet_{t_{1}}\\
~\label{eq:8th=00003D00003D00003D00003D00003D00003D000020diag}\nonumber 
\end{eqnarray}
which represents the case that the particle remains in the same run
state from $t_{5}$ to $t_{8}$, from $t_{3}$ to $t_{4}$, from $t_{1}$
to $t_{2}$, but changes to another run states at least twice, first
at some time between $t_{3}$ and $t_{2}$, and then again between
$t_{5}$ and $t_{4}$. Invoking the Markovian property, we see that
the probability for this diagram $\mathbb{P}\left(diag\right)$ can
be calculated ``segment by segment'': first there is a a factor
$\gamma_{T}/\left(\gamma_{R}+\gamma_{T}\right)$ for the probability
that particle is in a run state at $t_{1}$; then moving to the left,
each solid segment from $t_{k}$ to $t_{k-1}$ contributes a factor
$e^{-\gamma_{R}\left(t_{k}-t_{k-1}\right)}$, being the probability
that the particle remains in the same run state during the time between
$t_{k}$ to $t_{k-1}$; each blank from $t_{2k}$ to $t_{2k+1}$ contributes
\begin{eqnarray}
 &  & P\left(t_{2k+1},t_{2k}\right)=\frac{\gamma_{T}}{\gamma_{R}+\gamma_{T}}+\nonumber \\
 &  & +\frac{\gamma_{R}}{\gamma_{R}+\gamma_{T}}e^{-\left(\gamma_{R}+\gamma_{T}\right)\left(t_{2k+1}-t_{2k}\right)}-e^{-\gamma_{R}\left(t_{2k+1}-t_{2k}\right)}~~~~~,
\end{eqnarray}
which is the probability that the particle is in different run states
at $t_{2k}$ and $t_{2k+1}$. Note that $t_{2k}$ always connects
to $t_{2k-1}$. 

The conditioned expectation can be calculated from the moments of
the active velocity, as the condition requires the active velocity
to remain the same on a segment, and the expectation is taken only
over the initial choice of the active velocity. Thus, each solid segment
passing $2k$ points contributes a factor:

\begin{equation}
M_{d}^{k}=\frac{\Gamma\left(1/2+k\right)\Gamma\left(d/2\right)}{\sqrt{\pi}\Gamma\left(d/2+k\right)}.
\end{equation}

Therefore, the contribution from the diagram Eq. \ref{eq:8th=00003D00003D00003D00003D00003D00003D000020diag}
to the correlation function of order eight then amounts to: 
\begin{eqnarray}
pe^{-\gamma_{R}\sum_{k}\left(t_{2k}-t_{2k-1}\right)} & \times & P\left(t_{5},t_{4}\right)P\left(t_{3},t_{2}\right)\times\nonumber \\
 & \times & e^{-\gamma_{R}\left(t_{7}-t_{6}\right)}M_{d}^{2}\left(M_{d}^{1}\right)^{2}.
\end{eqnarray}

Using Eq. \ref{eq:MomentsGeneral}, (the Laplace transform of) the
moments can be calculated from the correlation function. Furthermore,
by breaking the diagram from the leftmost blank, one may obtain a
Volterra difference equation that can be programmed to recursively
calculate the moments. Define

\begin{equation}
L^{l}\left(\xi\right)=\frac{\mathscr{L}\left(\left\langle x^{2l}\right\rangle \right)\left(\xi\right)\xi\left(2lb+\xi\right)\left(\gamma_{R}+\gamma_{T}\right)}{\left(2l\right)!v^{2l}\gamma_{T}},\label{equdef}
\end{equation}
one can obtain: 
\begin{widetext}
\begin{align}
L^{l}\left(\xi\right)= & \sum_{k=1}^{l-1}\left(\prod_{m=2k+1}^{2l-1}\frac{1}{\gamma_{R}+mb+\xi}\right)g_{k}L^{k}\left(\xi\right)M_{d}^{l-k}+\prod_{m=+1}^{2l-1}\frac{1}{\gamma_{R}+mb+\xi}M_{d}^{l}.\label{eq:volterra=00003D00003D00003D00003D00003D00003D000020deq}
\end{align}
\end{widetext}

where: 
\begin{equation}
g_{k}=\frac{\gamma_{T}\gamma_{R}}{\left(\gamma_{R}+\gamma_{T}+2kb+\xi\right)\left(\gamma_{R}+2kb+\xi\right)\left(2kb+\xi\right)},\label{eq:gk}
\end{equation}
which comes directly from the Laplace transform of $P$.

In the special case $d=1$, or $d=2$ and $\gamma_{T}\rightarrow\infty$
(the zero tumble time limit), the exact steady state distribution
can be obtained using hypergeometric function. In the case $b=0$
(the free space limit), the exact Fourier-Laplace transform of the
time-dependent distribution can be calculated and agrees with \citep{2013_AveragedRunandTumbleWalks}.

\section{Active Brownian run-and-tumble particles}

\label{sec:Active-Brownian-run}

\subsection{Equation of motion}

\label{subsec:Equation-of-motion}

We are now ready to consider the ABRTP, the main topic of this paper.
The equation of motion is, formally:

\begin{equation}
\dot{x}\left(t\right)=-bx\left(t\right)+vF\left(t,K\left(t\right)\right),\label{fig:The-definition-of-K}
\end{equation}
$K\left(t\right)$ is a two state Markov process: 
\begin{equation}
R\underset{\gamma_{T}}{\overset{\gamma_{R}}{\rightleftharpoons}}T,
\end{equation}
where we use $R$ and $T$ to represent the run and tumble state,
respectively. The rates $\gamma_{R}$ and $\gamma_{T}$, both considered
constants.$F$ is defined as: 
\begin{equation}
F\left(t,T\right)=0,
\end{equation}
\begin{equation}
F\left(t,R\right)=\cos\theta\left(t\right),
\end{equation}
\begin{equation}
\dot{\theta}\left(t\right)=\sqrt{2D}\eta\left(t\right).
\end{equation}
Furthermore, each time $K$ enters $R$ from $T$, the value of $\theta$
is reset randomly.

Similar to both the RTP and the ABP, Eq. \ref{fig:The-definition-of-K}
implies $\left|x\left(t\right)\right|<v/b$ unless $\left|x\left(0\right)\right|>v/b$
and $t$ is small since since $F\left(t,K\right)\leqslant1$. This
can be proved by taking the absolute value of Eq. \ref{eq:SolutionGeneralODE},
or intuitively, this can also be expected from the equation of motion
Eq. \ref{subsec:Active-Brownian-particles-1}, since the particle
can move outward only when $0<x<v/b,$ and the closer to the boundary,
the slower its maximum possible velocity.

\subsection{Diagram laws}

\label{subsec:Diagram-laws}

The argument is very similar to the RTP, except now the conditioned
expectation$\left\langle \cdot|diag\right\rangle $ is given by averaging
over a ABP run state, which for simplicity we write as:
\begin{equation}
G\left(t_{j},...t_{k}\right)=v^{j-k+1}\left\langle \cos\theta\left(t_{j}\right)...\cos\theta\left(t_{k+1}\right)\cos\theta\left(t_{k}\right)|diag\right\rangle .
\end{equation}
 This is very similar to the correlation function for the ABP, except
that now,

\begin{equation}
\theta\left(t\right)=\theta\left(t_{k}\right)+\sqrt{2D}\int_{t_{k}}^{t_{j}}\eta\left(s\right)ds.
\end{equation}
and the underlying random variable is not simply the Gaussian white
noise $\eta$, but also $\theta\left(t_{k}\right)$. Although $\theta\left(t_{k}\right)$
is (almost surely) not (the angle of) the active velocity chosen at
the start of the run state, it does nevertheless have the same distribution.

To calculate the average, we again convert the cosine into exponential: 
\begin{widetext}
\begin{align}
G\left(t_{j},...t_{k}\right)= & \left(\frac{v}{2}\right)^{j-k+1}\sum_{a_{n}=\pm1}\left\langle e^{i\left(\sum_{n}a_{n}\right)\theta\left(t_{k}\right)}e^{ia_{n}\sqrt{2D}\sum_{m=1}^{j-k}\int_{t_{k}}^{t_{k+m}}\eta\left(s\right)ds}|diag\right\rangle 
\end{align}
\end{widetext}

Since $\theta\left(t_{k}\right)$ are independent from $\eta\left(s\right)$
for $s>t_{k}$, the average can be broken into the product of the
two exponential. The first average over $\left\langle e^{i\left(\sum_{n}a_{n}\right)\theta\left(t_{k}\right)}\right\rangle $
is non-zero only if $\sum_{n}a_{n}=0$, and the second is similar
to that in Subsection $\ref{subsec:Active-Brownian-particles-1}$,
thus we have: 
\begin{widetext}
\begin{eqnarray}
 &  & G\left(t_{j},...t_{k}\right)=\left(\frac{v}{2}\right)^{j-k+1}\times\sum_{a_{n}=\pm1,\sum_{n=k}^{j}a_{n}=0}\!\!\!\!\!\!\!\!\!\!\!\!\!\!e^{-D\sum_{m=k}^{j-1}\left(t_{m+1}-t_{m}\right)\left(\sum_{n=m+1}^{j}a_{n}\right)^{2}}~~~~~~
\end{eqnarray}
\end{widetext}

This form is ready for the Laplace transform. Together with the probability
of remaining in the same run state during the time $e^{-\gamma_{R}\left(t_{j}-t_{k}\right)}$,
and from the harmonic potential $e^{-b\sum_{m=k}^{j-1}m\left(t_{m+1}-t_{m}\right)}$,
the contribution from this run state to the Laplace transform of the
moment is (apart from the factor $\left(v/2\right)^{j-k+1}$):
\begin{widetext}
\begin{equation}
Q\left(j,k,\xi\right)=\sum_{a_{n}=\pm1,\sum_{n=k}^{j}a_{n}=0}\prod_{m=k}^{j-1}\frac{1}{\gamma_{R}+bm+D\left(\sum_{n=m+1}^{j}a_{n}\right)^{2}+\xi}.\label{eq:def=00003D00003D00003D00003D00003D00003D000020Q}
\end{equation}
\end{widetext}

The resulting diagram law is very similar to the RTP one, noticing
that $\sum_{n}a_{n}=0$ and $a_{n}=\pm1$ implies that the overall
contribution from one diagram will be nonzero only when every solid
line passes through even numbers of vertices, which in turn, implies
that every $2k$ vertices must connect to $2k-1$ vertices.

Thus the diagram law to calculate the Laplace transform of the $2l$
moment:

1) draw $2l$ vertices on a line, and connect vertex $2k$ to vertex
$2k-1$ for all $k\leqslant l$

2) connect vertex $2k-1$ to vertex $2k-2$ for some $k$

3) for every solid segment from vertex $k$ to vertex $j$, a factor
$Q\left(j,k,\xi\right)$; for every blank between vertex $2k+1$ and
vertex $2k$, factor $g_{k}$ as defined in Eq. \ref{eq:gk}; finally
there is an additional factor that is the same for every diagram:

\begin{equation}
\frac{\left(2l\right)!\gamma_{T}v^{2l}}{2^{2l}\xi\left(2lb+\xi\right)\left(\gamma_{R}+\gamma_{T}\right)}
\end{equation}
multiply them together to obtain the contribution from this diagram.

4) sum over all possibilities in 2) to obtain the $2l$ moment.

\subsection{Volterra difference equation}

\label{subsec:Recursive-relation}

Similar to the RTP problem, from the diagram laws, we may derive a
programmable Volterra difference equation that avoids calculating
all $2^{l-1}$ diagrams for $2l$ moment and considerably simplifies
the calculation. By breaking any diagram from the leftmost blank,
if one sets 
\begin{equation}
N^{l}\left(\xi\right)=\frac{\mathscr{L}\left(\left\langle x^{2l}\right\rangle \right)\left(\xi\right)2^{2l}\xi\left(2lb+\xi\right)\left(\gamma_{R}+\gamma_{T}\right)}{\left(2l\right)!\gamma_{T}v^{2l}},
\end{equation}
one can obtain: 
\begin{equation}
N^{l}\left(\xi\right)=\sum_{k=0}^{l-1}Q\left(2l,2k+1,\xi\right)g_{k}N^{k}\left(\xi\right),\label{eq:Vdeq=00003D00003D00003D00003D00003D00003D000020ABRTP}
\end{equation}
where it is understood that $g_{0}=N^{0}=1$.

Now we see how the ABRTP includes both the RTP and the ABP as the
special limit:

1) when $\gamma_{R}\rightarrow0$, $g_{k}\rightarrow0$ for all $k>0$,
and thus the only surviving term is $N^{l}\left(\xi\right)=Q\left(2l,1,\xi\right)$,
which can be shown to be the moments of the ABP by construction of
$Q$. Physically, $\gamma_{R}\rightarrow0$ means the particle will
remain in one run state indefinitely, and since ABRTP in the run state
behaves the same as the ABP, the moments will be the same.

2) when $D\rightarrow0$, $Q$ will reduce into: 
\begin{widetext}
\begin{equation}
Q\left(2l,2k+1,\xi\right)=\sum_{a_{n}=\pm1,\sum_{n=2k+1}^{2l}a_{n}=0}~~\prod_{m=2k+1}^{2l-1}\frac{1}{\gamma_{R}+bm+\xi},
\end{equation}
\end{widetext}

which is very similar to what appears in the Volterra difference equation
for the RTP Eq. \ref{eq:volterra=00003D00003D00003D00003D00003D00003D000020deq}.
Although $a_{n}$ does not directly appear in the equation, the restrain
$\sum_{n=2k+1}^{2l}a_{n}=0$ are still effective, and we need to count
all possible ways of assigning the $a_{n}$ to $\pm1$ under such
constrain. In other word: 
\begin{equation}
Q\left(2l,2k+1,\xi\right)=S\left(2l-2k\right)\!\!\!\prod_{m=2k+1}^{2l-1}\frac{1}{\gamma_{R}+bm+\xi},
\end{equation}
where $S\left(2m\right)$ is the numbers of possible ways of assigning
the $a_{n}$ to $\pm1$ such that $\sum_{n=1}^{2m}a_{n}=0$. This
differs from Eq. \ref{eq:volterra=00003D00003D00003D00003D00003D00003D000020deq}
by trading the dimension depending factor $M_{2}^{k}$ for a combinatorial
problem of $S$. 

To see that they are indeed equivalent, we note that value of $S\left(2k\right)$
can be calculated as the number of ways to place $k$ minus signs
to $2k$ possible locations in total, and therefore:
\begin{equation}
S\left(2k\right)=\left(\begin{array}{c}
2k\\
k
\end{array}\right)=\frac{\left(2k\right)!}{\left(k!\right)^{2}}.
\end{equation}
With the identity 
\begin{equation}
\frac{\left(2k\right)!}{2^{2k}k!}=\frac{\Gamma\left(k+\frac{1}{2}\right)}{\sqrt{\pi}},
\end{equation}
we see that
\begin{equation}
\frac{S\left(2k\right)}{2^{2k}}=M_{2}^{k},
\end{equation}
The identity itself can be proven by induction. For $k=0$ it is trivial.
Assuming it is true for $k$, then for $k+1$ we have:
\begin{equation}
\frac{\left(2k+2\right)!}{2^{2k+2}\left(k+1\right)!}=\frac{\left(2k+1\right)}{2}\frac{\Gamma\left(k+\frac{1}{2}\right)}{\sqrt{\pi}}=\frac{\Gamma\left(k+1+\frac{1}{2}\right)}{\sqrt{\pi}}.
\end{equation}
Therefore we have seen that the factor $S$ is indeed equivalent to
$M_{2}^{k}$. Therefore, we see that in the limit $D=0$, the ABRTP
model indeed reduce to the RTP model.

\subsection{Recursive relation for $Q$ }

\label{subsec:Recursive-relation-for}

Although $Q$ has an explicit form in Eq. \ref{eq:def=00003D00003D00003D00003D00003D00003D000020Q},
it involves selecting from $2^{j-k}$ cases and summation. Fortunately
there is a recursive relation that could simplify the calculation.

Notice that in Eq. \ref{eq:def=00003D00003D00003D00003D00003D00003D000020Q}
, term $a_{k}$does not formally appear, it only affect $Q$ by requiring
the summation of the rest of the $a_{n}$ to be $\pm1$. Thus define: 
\begin{widetext}
\begin{equation}
\hat{Q}\left(j,k,\xi,c\right)=\sum_{a_{n}=\pm1,\sum_{n=k+1}^{j}a_{n}=c}~~~\prod_{m=k}^{j-1}\frac{1}{\gamma_{R}+bm+D\left(\sum_{n=m+1}^{j}a_{n}\right)^{2}+\xi}\label{eq:Qhat=00003D00003D00003D00003D00003D00003D000020def}
\end{equation}
\end{widetext}

we have: 
\begin{equation}
Q\left(j,k,\xi\right)=\hat{Q}\left(j,k,\xi,1\right)+\hat{Q}\left(j,k,\xi,-1\right).
\end{equation}

On the other hand, applying $\sum_{n=k+1}^{j}a_{n}=c$ to Eq. \ref{eq:Qhat=00003D00003D00003D00003D00003D00003D000020def},
we notice that term $a_{k+1}$ will no longer appear, and thus we
have the recursive relation: 
\begin{equation}
\hat{Q}\left(j,k,\xi,c\right)=\frac{\hat{Q}\left(j,k+1,\xi,c-1\right)+\hat{Q}\left(j,k+1,\xi,c+1\right)}{\gamma_{R}+bk+Dc^{2}+\xi}.\label{eq:Qhat=00003D00003D00003D00003D00003D00003D000020recur}
\end{equation}

To complete, we need proper initial condition. The first is obvious
that 
\begin{equation}
\hat{Q}\left(j,j-1,\xi,\pm1\right)=\frac{1}{\gamma_{R}+b\left(j-1\right)+D+\xi}.
\end{equation}
Another condition is that, since $\left|a_{n}\right|=1$, if $\left|c\right|>j-k$,
then constrain $\sum_{n=k+1}^{j}a_{n}=c$ is impossible, thus:

\begin{equation}
\hat{Q}\left(j,k,\xi,\left|c\right|>j-k\right)=0.
\end{equation}

These results can be used to recursively calculate $\hat{Q}$ and
therefore $Q$, since on the LHS of Eq. \ref{eq:Qhat=00003D00003D00003D00003D00003D00003D000020recur},
$j-k$ always decreases by one and thus eventually can be reduced
to $j-k=1$, which can be obtained from the initial conditions.

\section{Results}

\label{sec:Results}

\subsection{Free space: an experimental cross-check }

\label{subsec:Free-space}

The model successfully captures experimental results. Here we use
data obtained from the run-and-tumble motion of an immune cell (BV2
cell line) in 2D free space. The experimental details of the mechanotaxis
in microglial migration have been described in Ref. \citep{2023_RunandTumbleDynamicsandMechanotaxisDiscoveredinMicroglialMigration}.
While our calculations were done for the case of a harmonic trap,
it is straightforward to obtain the corresponding results for free
space, simply by setting $b=0$ in our formulas. Thus our theoretical
calculation can be easily checked against the free space experimental
data. The only difference is that in free space there is no equilibrium
and we must consider the time-depending problem. 

The specific dataset, that we will use for comparison with theoretical
predictions, consists of $N=244$ cell migration tracks of on a flat
collagen substrate. The tracks are sampled at a rate of 2 frames/min
and last up to 500 min. The relevant parameters extracted from the
experimental migration tracks include: the state transition rates
$\gamma_{R}=0.58\min^{-1}$,$\gamma_{T}=0.091\min^{-1}$, the speed
of the runs $v_{e}=6.7\mu{\rm {m}\cdot\min^{-1}}$, and the angular
diffusion during the run state $D=0.67\min^{-1}$. 

The mean squared displacement (MSD) of the migrating cell as a function
of time are first compared to predictions of the ABRTP model, which
can be cast into the form: 
\begin{equation}
\left\langle r^{2}\right\rangle =\frac{2v^{2}\gamma_{T}}{\gamma_{R}+\gamma_{T}}\left(\frac{e^{-\left(D+\gamma_{R}\right)t}-1}{\left(D+\gamma_{R}\right)^{2}}+\frac{1}{D+\gamma_{R}}t\right).
\end{equation}
The above expression entails a ballistic scaling of the MSD at short
time scales ($\left\langle r^{2}\right\rangle \propto t^{2}$) and
a diffusive scaling ($\propto t$) at long time scales, which both
capture qualitatively the experimental trend. However, as similar
behaviors are also predicted by the RTP models, one must seek for
other measures to showcase the better applicability of the ABRTP model.

For this we again employ the expectation of Chebyshev polynomials.
The migration tracks are segmented into 45 min samples ($N=1496$).
$x$ and $y$ components of the tracks are then separately processed
as two different datasets. All segments are aligned to start at the
origin and rescaled to be within $\left[-1,1\right]$ by the largest
theoretically possible traveling distance, \textsl{i.e}., $v$ times
the segment length 45min, so that the experimental migration tracks
mimic the conditions used in theoretical calculations. Finally, the
theoretical expectations of the Chebyshev polynomials are calculated
by the method outlined in Section \ref{sec:Active-Brownian-run},
with the reported parameters $\gamma_{R}$, $\gamma_{T}$, $D$ \citep{2023_RunandTumbleDynamicsandMechanotaxisDiscoveredinMicroglialMigration}.
As for the active velocity $v$, the measured speed $v_{e}$ is interpreted
as the average speed over the time between successive observations,
\textsl{i.e.}, $0.5\min$. We choose the corresponding model parameter
of active velocity $v$ for the ABRTP model as $v_{m}=7.9\mu{\rm {m}\cdot\min^{-1}}$,
such that $\left\langle x\left(0.5\min\right)|\theta\left(0\right)=0\right\rangle =v_{e}*0.5\min$.
In other word, the $v_{m}$ is so chosen such that the averaged velocity
over $0.5\min$ is roughly $v_{e}$. For comparison, we also present
the calculations using the RTP model with both $v_{e}$ and $v_{m}$.
As for the ABP model, it does not capture the long, dominating tumble
time with no active motion, and the results fail miserably and not
shown.

\begin{figure*}
\centering \includegraphics[width=0.8\textwidth]{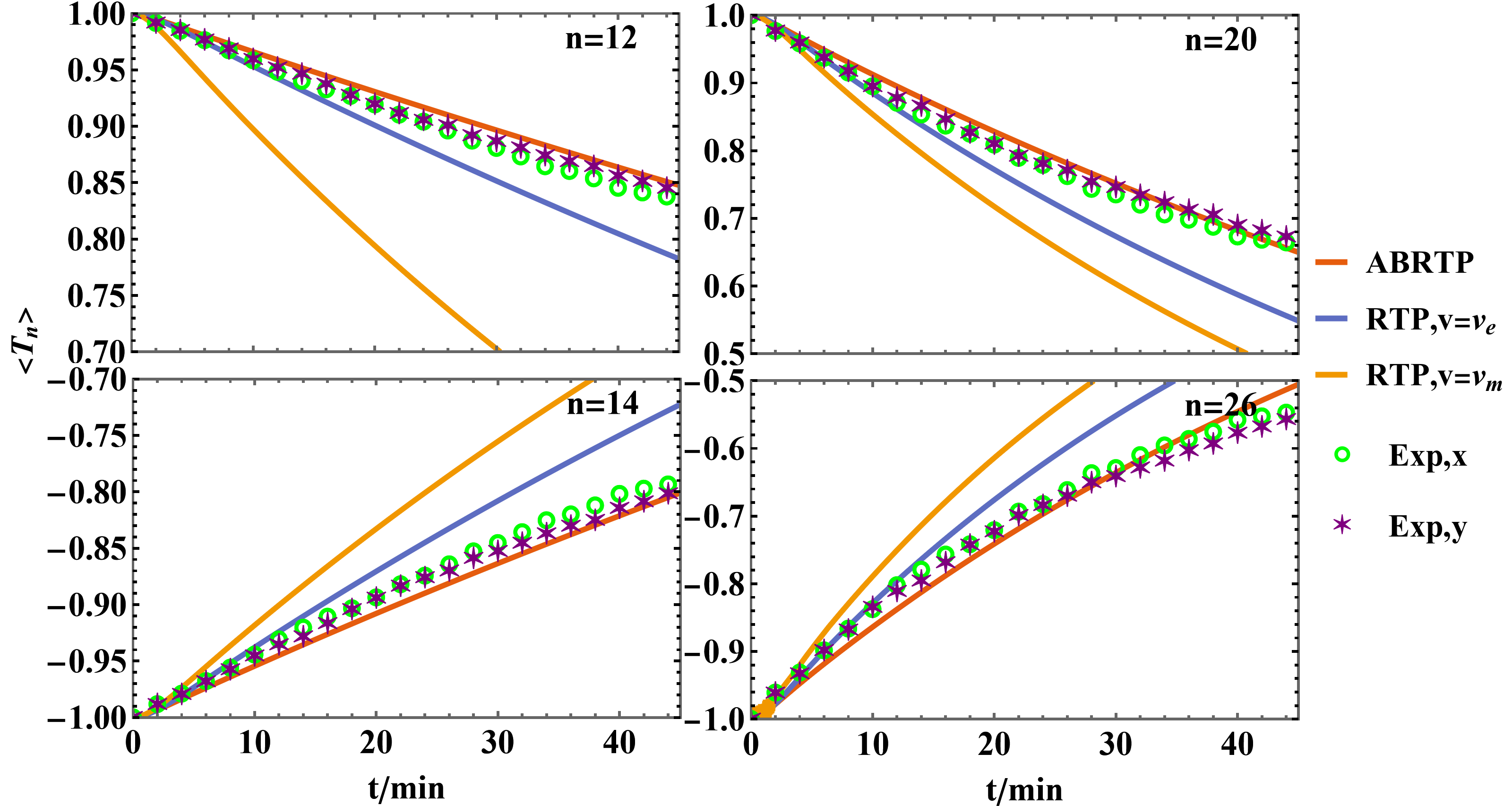}\caption{The time depending expectation of Chebyshev polynomials of the active
particles in free space. Dot: experimental data from the BV2 cells,
Exp $x$ means results from the $x$ components of the experimental
data, and Exp $y$ means the y components. See \citep{2023_RunandTumbleDynamicsandMechanotaxisDiscoveredinMicroglialMigration}
for the details of the experiments; line: analytic calculation from
Eq. \ref{eq:Vdeq=00003D00003D00003D00003D00003D00003D000020ABRTP},
with parameters $\gamma_{R}=0.58\min^{-1}$,$\gamma_{T}=0.091\min^{-1}$,$D=0.67\min^{-1}$
as reported in the same reference. As for the active velocity, $v_{e}=6.7\ \mu m/\min$
represents the averaged velocity of the run state in the interval
of $0.5\min$ as reported, $v_{m}=7.9\ \mu m/\min$ is the inferred
instantaneous velocity, or a model parameter that results in the experimentally
averaged velocity $v_{e}$ according to the ABP model. In the calculation
of the ABRTP model $v_{m}$ is used. It is obvious that the ABRTP
model captures the experimental results better than the RTP model.}
\label{fig:Experimental-check} 
\end{figure*}

Results of different models alongside with experimental data are presented
in Fig.~\ref{fig:Experimental-check}. The results from the $x$
and $y$ components of the data are compatible with each other, as
they correspond to an isotropic distribution. It is obvious that the
ABRTP model captures the experimental results well. On the other hand,
the RTP model fares less well and fails to reproduce the experimental
trends, though the RTP model with $v_{e}$ does a slightly better
job, especially at small times.

\subsection{Steady state in harmonic trap }

\label{subsec:Steady-state-in}

In this subsection we rescale the parameters so that $b=v=1$. A most
interesting and apparent aspect of the steady state of an active particle
in a harmonic trap is that they might cluster near the boundary, where
the potential energy is the highest. This is in obvious contrast to
the Boltzmann distribution that concentrates the particles near the
center. Specifically, the ABP will concentrate near, {\em but not
at}, the boundary, giving a smooth distribution \citep{2024_ActiveOscillatorRecurrenceRelationApproach},
whereas the distribution of the RTP, depending on the value of $\gamma_{R}$
and $\gamma_{T}$, could be singular at the boundary and/or the center.

\begin{figure*}
\centering \includegraphics[width=0.8\textwidth]{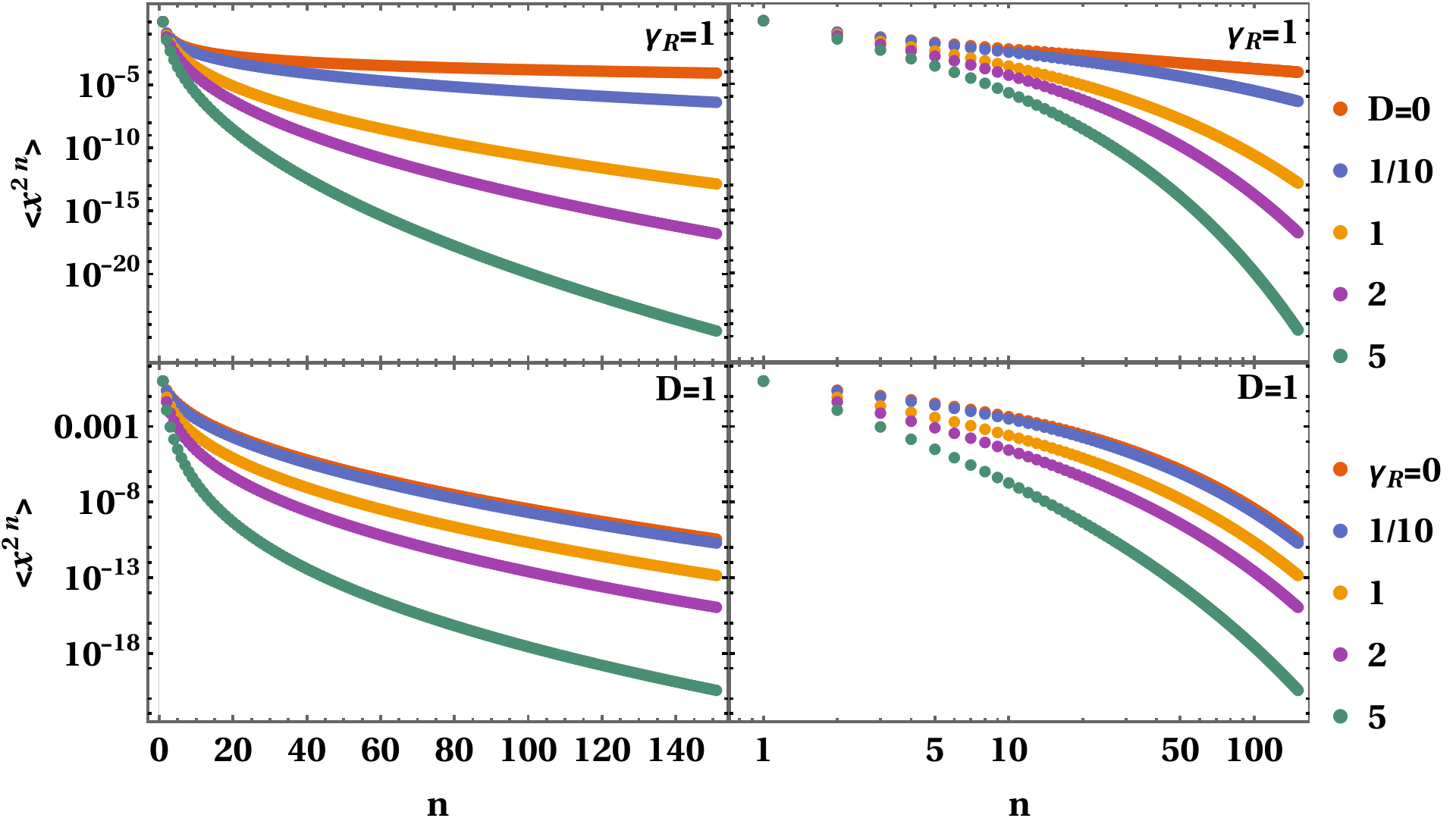}\caption{Log (left column) and log-log (right column) plots of the (rescaled)
moments as a function of their order, as calculated using Eq. \ref{eq:Vdeq=00003D00003D00003D00003D00003D00003D000020ABRTP}.
$\gamma_{T}=1$ in all cases, but its exact value has little effect
on the asymptotic behavior. Unlike the RTP that features an algebraic
scaling law, the ABRTP seems to exhibit an geometric scaling law,
and asymptotic to a line in the log plot.}
\label{fig:moment=00003D00003D00003D00003D00003D00003D000020scaling=00003D00003D00003D00003D00003D00003D000020law} 
\end{figure*}

The distribution near the boundary is directly related to the asymptotic
behavior of the higher order moments. In the case of ABRTP, from the
definition of $Q$ in Eq. \ref{eq:def=00003D00003D00003D00003D00003D00003D000020Q},
it is apparent that, for the ABRTP the steady state moments will be
smaller than both the RTP ($D=0$) and the ABP ($\gamma_{R}=0$),
since the denominator is larger. It might also be naively expected
that the high order moments will resemble the ABP more, since $\gamma_{R}$
contributes the same to all denominators whereas $D$ contributes
as $D\left(\sum_{n=m+1}^{j}a_{n}\right)^{2}$ and can be very large
when the term is in a long line segment in the diagram. These insights
are corroborated by detailed numerics: in contrast to the RTP, which
features a algebraic high order moments scaling law determined by
$\gamma_{R}$, the ABRTP, as long as $D>0$, appears to exhibit a
geometric scaling law according to the log plot, with the slope depending
mostly on $D$, as shown in Fig. \ref{fig:moment=00003D00003D00003D00003D00003D00003D000020scaling=00003D00003D00003D00003D00003D00003D000020law}.
Therefore we conclude that the ABRTP behaves similarly to the ABP
near the boundary, and thus might cluster somewhere near, but not
at, the boundary where the density vanishes.

On the other hand, the behavior near the center in the case of ABRTP
cannot be inferred from the asymptotic behavior of the higher order
moments alone. Indeed, the density near the center contributes little
to high order moments, and cannot be determined unless all the moments
are known exactly. Nevertheless, with some reasonable assumptions
it is still possible to obtain some valuable insights into its behavior.
In fact since the tumble state is the same for both models, we expect
the ABRTP to behave like RTP near the center, and thus a small $\gamma_{T}$
will produce a sharp peak or even singularity near the center, whereas
in the run state, a large value of $\gamma_{R}$ and $D$ will be
responsible for the frequent turning and a continuous peak near the
center.

\begin{figure*}
\centering \includegraphics[width=0.8\textwidth]{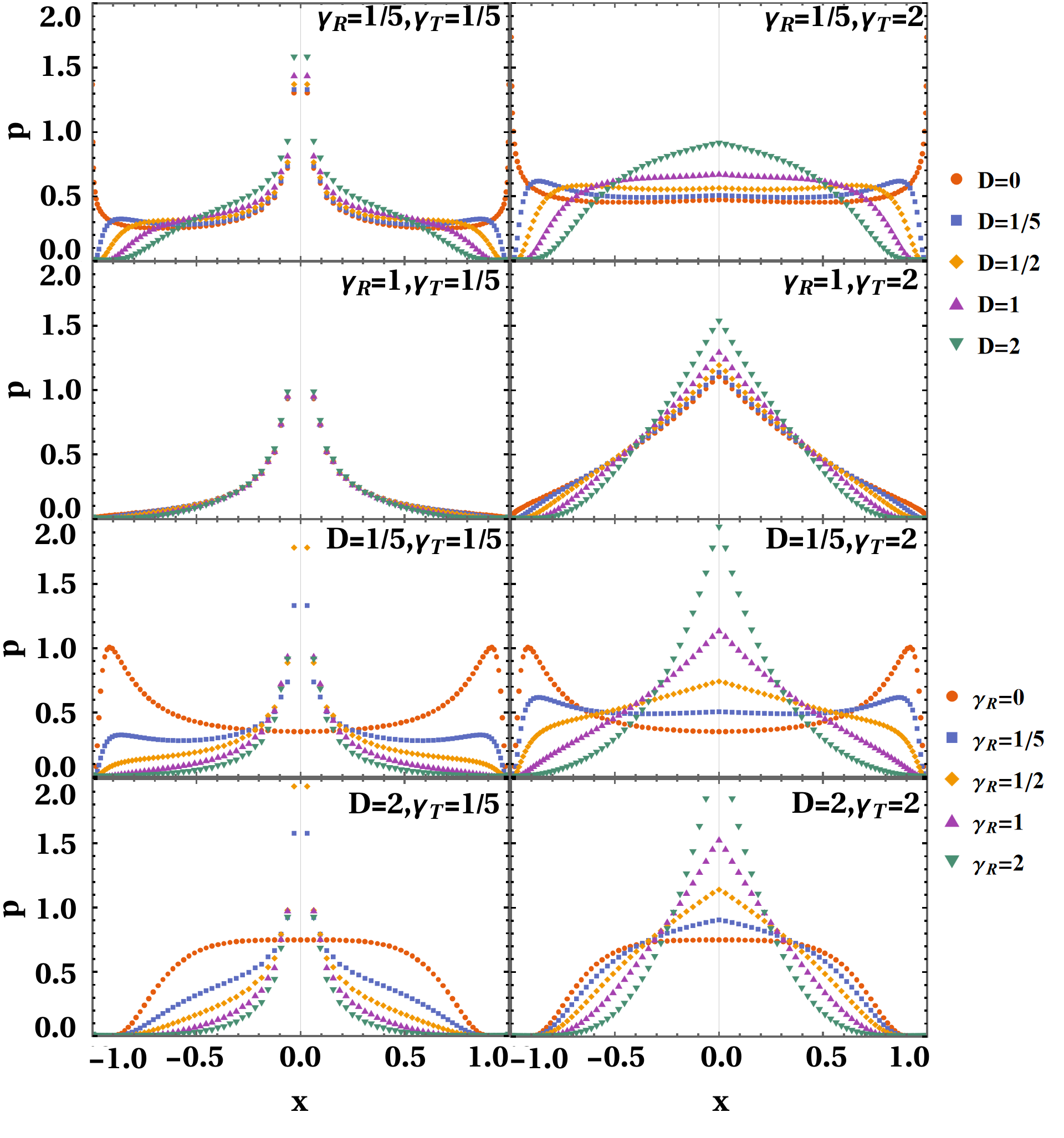}\caption{Steady state distributions inferred from the moments using methods
outlined in Section \ref{sec:Kac's-method-revisited}. The most obvious
feature of such distribution are the possibly singular peaks near
the center and/ or the boundary. At the boundary, the ABRTP behaves
mostly like the ABP that clutter near, but not at, the boundary, thus
a continuous peak near the boundary for small $\gamma_{R}$ and $D$.
At the center, the ABRTP behaves mostly like the RTP, with a possible
singular peak for small $\gamma_{T}$, and a continuous peak for large
$\gamma_{R}$ or $D$. Note that when $\gamma_{R}=0$, the particle
will be an ABP and never enters a tumble state, thus $\gamma_{T}$
has no effect on the distribution. }
\label{fig:ssd} 
\end{figure*}

To support these arguments and to provide a more vivid insight into
the character of the steady state, we attempt to infer the distribution
from the moments, using the method outlined in Section \ref{sec:Kac's-method-revisited}.
Examples of the resulting approximate distributions are shown in Fig.
\ref{fig:ssd}. The distribution appears to agree with the above qualitative
arguments, with a single trivial exception corresponding to $\gamma_{R}=0$.
In this case alone the ABRTP reduces to ABP, that never enters a tumble
state. $\gamma_{T}$ therefore does not matter, and small $\gamma_{T}$
does not result in a peak at the center as in the case of the RTP.

\section{Conclusion}

\label{sec:Conclusion}

The Kac method of directly integrating the equation of motion, that
has been to some extent overlooked in favor of the Fokker-Planck equation,
is in fact a very powerful methodological approach in studying certain
problems of statistical mechanics. By properly applying the Kac methodology,
we have been able to duplicate many of the existing theoretical results
regarding the RTP and the ABP, as well as obtain some new results.
This allows us to combine the two models together in order to model
the active particles better, resulting in the ABRTP, that combines
many features of both the RTP and the ABP. In free space, the MSD
shows a ballistic behavior $\left\langle r^{2}\right\rangle \propto t^{2}$
at short time scales and a diffusive behavior $\left\langle r^{2}\right\rangle \propto t$
at long time scales, similar to both the RTP and the ABP. In the presence
of a harmonic trap, the steady state near the boundary is similar
to the ABP, and near the center its steady state is similar to the
RTP near the center. 

The results that are obtainable from the Kac method are the moments,
not the full distribution. However, in many problems of statistical
physics we can often obtain the full distribution directly, while
its moments can be calculated from this distribution. Therefore our
methodology, moments -> statistical distribution, appears to be in
some sense {\em dual to the standard approach}, statistical distribution
-> moments, and allows us to explore some problems where other known
methodologies are easily applicable. In order to relate our results
to experiments or numerical simulations, we attempt successfully to
compare the expectation of the Chebyshev polynomials. To analyze the
distribution, we first approximate the expectation of a smooth function
by using the Chebyshev polynomials, and then proceed to an estimation
of the distribution. 

It is known that the Chebyshev polynomials already play a vital role
in the pseudo-spectra method of numerical solutions of PDE \citep{2013_ChebyshevandFourierSpectralMethodsSecondRevisedEdition}
and in the method of root finding \citep{2014_SolvingTranscendentalEquations}.
It is, however, less clear that they might also be important in the
statistical data processing. In this paper we show that they can be
applied when the parameters of the model can be obtained directly
from the experimental data. It remains to be seen if such methodology
can be adapted to the parameter estimation in the case where they
are unknown or not known exactly.

The next problem is encountered when one ventures into 3D. One advantage
of the Fokker-Planck equation is that diffusion on a sphere is relatively
easy to formulate and solve, while the Langevin equation for diffusion
on a sphere is much more difficult. 

Finally, it would be interesting to see more applications of the Kac
method in the case of general problems in stochastic phenomena. Some
obvious generalizations of the RTP and ABRTP might be possible, \textsl{e.g}.,
the RTP biased in one direction, the RTP with multiple run states
with different velocities, \textsl{etc}.. Non trivial, and not so
obvious, applications in other fields of study would certainly be
highly interesting.

\section{Acknowledgments}

FY ackowledges the support of the National Natural Science Foundation
of China (Grant No. 12090054 and 12325405). RP acknowledges the support
of the Key project of the National Natural Science Foundation of China
(NSFC) (Grant No. 12034019).

 \bibliographystyle{unsrt}
\addcontentsline{toc}{section}{\refname}\bibliography{lib}

\end{document}